\documentclass[letterpaper]{article} 
\usepackage{aaai2026}  
\usepackage{times}  
\usepackage{helvet}  
\usepackage{courier}  
\usepackage[hyphens]{url}  
\usepackage{graphicx} 
\usepackage{natbib}  
\usepackage{caption} 
\usepackage{algorithm}
\usepackage{algorithmic}
\usepackage{soul}  
\usepackage{color} 
\usepackage{xspace}
\usepackage{listings}
\usepackage{enumitem}
\newcommand{\eg}{{\it e.g.,\ }}

\newcommand{\ie}{{\it i.e.,\ }}

\usepackage{booktabs}
\definecolor{oxfordblue}{rgb}{0.0, 0.13, 0.28}
\definecolor{harvardcrimson}{rgb}{0.79, 0.0, 0.09}
\definecolor{dartmouthgreen}{rgb}{0.05, 0.5, 0.06}
\definecolor{princetonorange}{rgb}{1.0, 0.56, 0.0}
\definecolor{yaleblue}{rgb}{0.06, 0.3, 0.57}
\definecolor{usccardinal}{rgb}{0.6, 0.0, 0.0}
\definecolor{uclablue}{rgb}{0.33, 0.41, 0.58}
\definecolor{msugreen}{rgb}{0.09, 0.27, 0.23}
\definecolor{cornellred}{rgb}{0.7, 0.11, 0.11}
\definecolor{pomegranate}{RGB}{192, 57, 43}
\definecolor{anti-pomegranate}{RGB}{43,178,192}
\definecolor{alizarin}{RGB}{231, 76, 60}
\definecolor{anti-belize}{RGB}{185, 41, 56}
\definecolor{belize}{RGB}{41, 128, 185}
\definecolor{peter}{RGB}{52, 152, 219}
\definecolor{green}{RGB}{22, 160, 133}
\definecolor{anti-green}{RGB}{160,22,118}
\definecolor{turquoise}{RGB}{26, 188, 156}
\definecolor{pumpkin}{RGB}{211, 84, 0}
\definecolor{anti-pumpkin}{RGB}{0,22,211}
\definecolor{carrot}{RGB}{230, 126, 34}
\definecolor{wisteria}{RGB}{142, 68, 173}
\definecolor{anti-wisteria}{RGB}{99,173,68}
\definecolor{amethyst}{RGB}{155, 89, 182}
\definecolor{nephritis}{RGB}{39, 174, 96}
\definecolor{anti-nephritis}{RGB}{174,39,117}

\newcommand{\lxe}[1]{{\color{black} #1}}
\newcommand{\pzh}[1]{{\color{black} #1}}
\newcommand{\peng}[1]{{\color{black} #1}}

\newcommand{\lxeee}[1]{{\color{black} #1}}
\newcommand{\penguin}[1]{{\color{black} #1}}
\newcommand{\lxee}[1]{{\color{black} #1}}
\newcommand{\ms}[1]{{\color{black} #1}}
\newcommand{\revise}[1]{{\color{black} #1}}

\usepackage{newfloat}
\usepackage{listings}
\DeclareCaptionStyle{ruled}{labelfont=normalfont,labelsep=colon,strut=off} 
\floatstyle{ruled}
\newfloat{listing}{tb}{lst}{}
\floatname{listing}{Listing}
\title{Understanding Fortunetelling with Large Language Models \penguin{in China}: User Practices, Perceptions, and Impacts on Beliefs and Decisions}
\author{
    Xueer Lin\textsuperscript{\rm 1},
    Chenyu Li\textsuperscript{\rm 1},
    Shuai Ma\textsuperscript{\rm 3},
    Yuhan Lyu\textsuperscript{\rm 1},
    Zhenhui Peng\textsuperscript{\rm 1}\thanks{Corresponding author.}
}

\affiliations{
    \textsuperscript{\rm 1}Sun Yat-sen University, Zhuhai, China\\
    \textsuperscript{\rm 2}Aalto University, Helsinki, Finland\\
    linxer6@mail2.sysu.edu.cn,
    lichy235@mail2.sysu.edu.cn,
    shuai.ma@connect.ust.hk,
    lvyh26@mail2.sysu.edu.cn,
    pengzhh29@mail.sysu.edu.cn
}

\usepackage{bibentry}

\usepackage{booktabs}
\usepackage{multirow}
\usepackage{makecell}
\usepackage[table]{xcolor}
\usepackage{colortbl}
\usepackage{subcaption}
\usepackage{csquotes}

\ExplSyntaxOn
\cs_if_exist:NTF \autoref { }{
  \cs_new_protected:Npn \autoref #1 { \aaai_autoref:n {#1} }
}
\cs_new_protected:Npn \aaai_autoref:n #1 {
  \str_if_eq:eeTF { \str_range:nnn {#1}{1}{4} } {sec:} {Section~\ref{#1}}{
  \str_if_eq:eeTF { \str_range:nnn {#1}{1}{4} } {fig:} {Figure~\ref{#1}}{
  \str_if_eq:eeTF { \str_range:nnn {#1}{1}{4} } {tab:} {Table~\ref{#1}}{
  \str_if_eq:eeTF { \str_range:nnn {#1}{1}{4} } {alg:} {Algorithm~\ref{#1}}{
  \str_if_eq:eeTF { \str_range:nnn {#1}{1}{4} } {app:} {Appendix~\ref{#1}}{
  \str_if_eq:eeTF { \str_range:nnn {#1}{1}{3} } {eq:}  {Eq.~(\ref{#1})}{
    \ref{#1}
  }}}}}}
}
\ExplSyntaxOff

\begin{document}

\maketitle

\begin{abstract}
\revise{Fortunetelling is a cultural practice for navigating uncertainty, often associated with people’s beliefs and decisions.}
Fortunetelling with recent large language models (LLMs) introduces new opportunities and risks. 
This paper \penguin{conducts qualitative studies} to understand users' practices, perceptions, and impacts of LLM fortunetelling \penguin{in China}. 
We first analyze 1,045 posts on Chinese social media, yielding a comprehensive taxonomy of the diverse foretold topics (\eg career, romance), emotion reactions (\eg surprise, worry), and perceived credibility (\eg doubt, trust) of LLM fortunetelling. 
Then, we conduct interviews with 20 users of LLM fortunetelling.
The findings indicate that users treat LLM fortunetelling as a tool less for accurate prediction but more for emotional support. 
\revise{While the fortunetelling results rarely change users' initial beliefs or decisions, they are associated with subtle mindset shifts, with some users reporting small behavioral adjustments.}
We discuss implications for gaining benefits from LLM fortunetelling.
\end{abstract}

\ms{
\section{Introduction}

\penguin{Fortunetelling, which attempts to predict or reveal information about a person's future, character, or fate through symbolic systems, rituals, or intuition, has long been practiced in many cultures \cite{smith2019overview}}. 
\penguin{
Specifically, in Chinese culture focused in this paper, fortunetelling is traditionally facilitated by human masters, conducted via metaphysical systems like Bazi and Ziwei Doushu (Purple Star Astrology), and used for important decisions such as marriage matching and naming children. 
Recent advances in large language models (LLMs) have boost a new accessible form of fortunetelling, \ie \textit{LLM fortunetelling}, actively discussed on social media platforms.} 
People increasingly turn to LLMs to seek predictions about major life events (\eg romantic prospects, academic success) and share these experiences on social media. For instance, on RedNote --- a popular Chinese platform --- users often prompt LLMs to act as fortunetelling masters by providing personal details such as birth date, time, and zodiac, and then discuss the predictions with others \revise{(\autoref{fig:three_figs})}. 

This phenomenon introduces both opportunities and risks. On the one hand, LLM fortunetelling can provide emotional support, stimulate self-reflection, and guide daily decision-making \cite{tan2022positive}. 
On the other hand, prior ICWSM work has already raised concerns about LLMs in sensitive advice settings (\eg relationship advice) \cite{hou2024chatgpt}.
LLM fortunetelling inherits challenges of LLMs such as hallucinations \cite{meier2022fortunetelling} and may encourage over-reliance, potentially leading to poor financial, academic, or career-related decisions \cite{grall2015fortune}. 
This ``double-edged sword'' calls for a deeper understanding of how people experience, perceive, and respond to LLM fortunetelling, so that its benefits can be harnessed while mitigating its risks. 

\captionsetup[subfigure]{justification=centering, font=small}
\begin{figure*}[ht!]
    \centering
    \begin{subfigure}[b]{0.32\textwidth}
        \includegraphics[width=\linewidth]{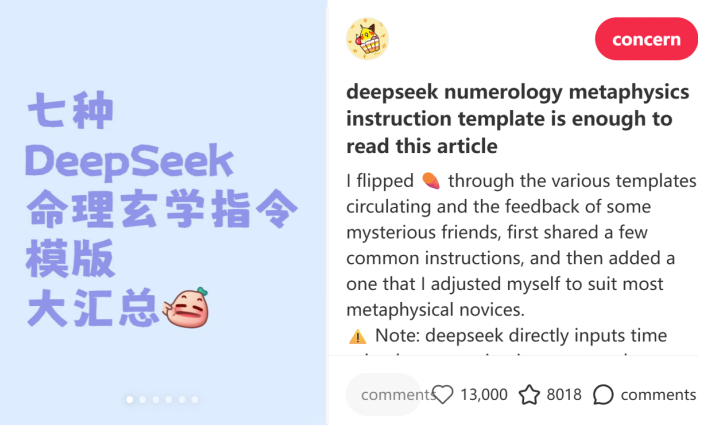}
        \caption{Sharing LLM Fortunetelling Prompt}
    \end{subfigure}
    \hfill
    \begin{subfigure}[b]{0.32\textwidth}
        \includegraphics[width=\linewidth]{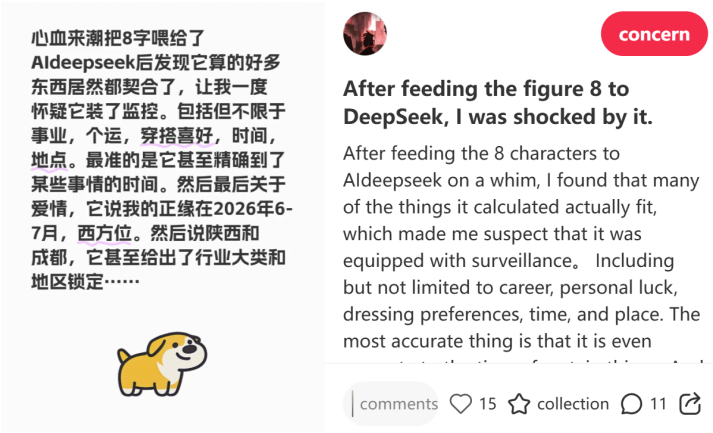}
        \caption{Sharing LLM Fortunetelling Experience}
    \end{subfigure}
    \hfill
    \begin{subfigure}[b]{0.32\textwidth}
        \includegraphics[width=\linewidth]{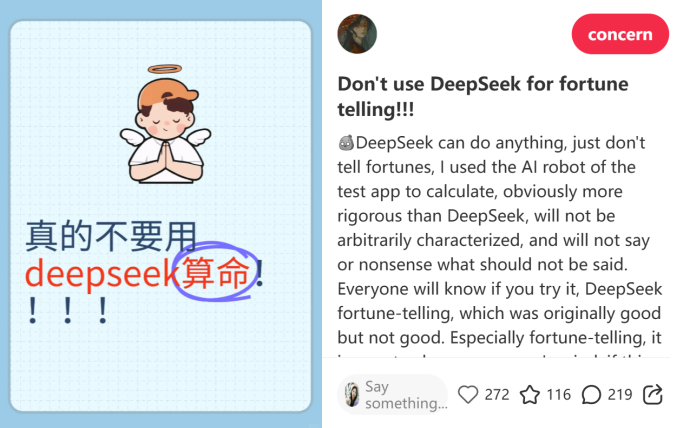}
        \caption{Discussing LLM Fortunetelling }
    \end{subfigure}
    
    \caption{Three topics of posts. Post content is translated from Chinese to English using the embedded translator of the browser.}
    \label{fig:three_figs}
\end{figure*}
However, the use of LLMs for fortunetelling —particularly how users engage with its predictions — remains largely unexplored. 
First, although LLM fortunetelling is often treated as a playful use case, prior research suggests that casual interactions with LLMs can acquire personal meaning and shape users' beliefs \cite{10.1145/3706598.3713408}.
Second, LLM fortunetelling is increasingly invoked for decision-making support. Yet, existing studies on AI-assisted decision-making have largely centered on expert or objective domains such as clinical diagnosis \cite{tsai2021exploring} or financial forecasting \cite{li2025text}. 
Unlike these contexts, fortunetelling concerns users’ own lives and choices, making AI-generated responses personally meaningful. Investigating these practices can therefore deepen our understanding of AI-mediated decision-making and belief formation in self-referential contexts.

To address these gaps, we investigate the following research questions \penguin{in the Chinese context}:}
\textbf{RQ1)}: \peng{What are the user practices of LLM fortunetelling regarding the foretold topics, prompts, and motivations?}
\textbf{RQ2)}: \lxe{What are the user perceptions of LLM fortunetelling?}
\textbf{RQ3)}: \peng{How would the users' beliefs and decisions on the foretold topics be affected by LLM fortunetelling?}


To answer these questions, \pzh{we first analyzed 1,045 posts about fortunetelling with LLMs from RedNote and Weibo -- two representative social media platforms in China, providing insights from a large and diverse user base to address RQ1 and RQ2. 
\peng{Then, we conducted interviews with 20 users of LLM fortunetelling to gain in-depth qualitative answers to each RQ.
\penguin{These two qualitative} studies yield a comprehensive taxonomy of user practices and perceptions of LLM fortunetelling. 
We found that people adopt well-structured prompts to consult LLMs on both traditional topics like Bazi (general fortune) and issues in modern society like pursuing postgraduate studies and locating lost property. 
\lxeee{As for user perceptions of and responses to LLM fortunetelling, our findings suggest the interplay of motivated reasoning, confirmation bias, and the Barnum effect, where users tend to interpret vague or personalized outputs in ways that reinforce their prior beliefs, expectations, and emotional needs \cite{nickerson1998confirmation, kunda1990case,huizi2022understanding}.}
Most users were positive towards LLM fortunetelling, feeling surprised when predictions coincidentally matched reality and joking about the implausible results. 
Besides, participants in the interview study reported that they turned to LLM fortunetelling for both decision support and emotional comfort. 
When interacting with the LLM, they proactively managed risks and
filtered the fortunetelling results. 
}}

\pzh{
In summary, this work contributes to a comprehensive understanding of \penguin{LLM fortunetelling's practices, user perceptions, and impacts on users' beliefs and decisions.}
The findings \penguin{reveal \lxeee{motivated reasoning, confirmation bias, and the Barnum effect} in LLM fortunetelling, suggesting} a new perspective on the role of LLM, via fortunetelling activities, in emotion regulation and decision support. 
Finally, this work highlights the potential implications of LLM fortunetelling on social media, where LLM-generated narratives shape how uncertainty is publicly expressed and managed, with subtle consequences over time.
}
\section{Related Work}
\subsection{\pzh{Fortunetelling as a Cultural Practice}}
Fortunetelling exists in diverse cultural contexts, from Eastern traditions like Bazi and Ziwei Doushu to Western customs like tarot cards and astrology. 
\lxe{For example, in China, people consult Bazi not only to form beliefs about their life trajectory and future potential, but also to guide concrete decisions such as choosing an auspicious date for marriage.}
Psychological and consumer behavior research suggests that positive fortunetelling experiences can enhance subjective well-being and increase risk-taking \cite{tan2022positive}, while dependence on fortunetelling may negatively impact decision-making and mental health \cite{skryabin2020addiction}. 

The rise of large language models (LLMs) has made AI-mediated fortunetelling more accessible. Related research has explored models optimized for divination-like outputs \cite{cheng2025beat}, while human-computer interaction researchers have developed multimodal conversational agents inspired by shamanic symbolism to study symbolic AI interaction \cite{10.1145/3706598.3714297}.
Recent studies on digital spirituality further show that social media enables the creation and circulation of symbolic meaning. For example, astrology content on social media shapes users’ self-concepts and relationships, while platforms such as TikTok have been described as ``sacred technologies'' that reinforce spiritual narratives \cite{doi:10.1177/14614448251344588}.
\pzh{These related studies collectively indicate the trends of using technology to facilitate traditional fortunetelling practices.}
However, there is a lack of systematic understanding of users' experiences with the advanced LLM fortunetelling. 
Our research fills this gap by providing insights into how users turn to LLMs to foretell everyday issues, how they feel about LLM fortunetelling, and how their beliefs and decisions are affected. 

\subsection{LLM in Playful Contexts, for Emotional Support, and Its Impact on Human Beliefs}
Previous research has integrated LLMs into informal everyday contexts, indicating that even seemingly playful interactions with LLMs can have unexpected meanings and consequences. 
\lxe{For example, research on BleacherBot \cite{10.1145/3706598.3714178} shows that AI can serve as a companion while watching sports. Because it is not a real person, users feel more comfortable without the usual social pressures of watching games with others.}
These findings collectively highlight the characteristics of LLM use: interactions may appear playful, but can subtly influence users' emotions, reflections, and decisions. 
\lxe{More recent work further illustrates the potential of LLM to provide emotional support and affect user beliefs.
For example, research on ChatLab \cite{10.1145/3706598.3713453} shows how users construct personalized LLM-powered chatbot personas to confront stressors, foster emotional reliance, and engage in self-reflection.}
\lxee{Empirical evidence further illustrates how LLM behaviors can shape belief formation. 
The research conducted by \citet{10.1145/3706598.3713408} finds that deceptive AI-generated explanations are more persuasive than accurate ones and could significantly increase belief in false news while eroding belief in true news.
}
Our work extends this understanding by investigating how seemingly playful interactions shape users' emotions and beliefs in the context of LLM fortunetelling.


\subsection{AI-Powered Decision-Making Support} 

\lxe{A growing body of research has explored AI-powered decision-making, mostly in domains where decisions are made by experts}, such as clinical diagnosis \cite{tsai2021exploring} and income forecasting \cite{li2025text}, where professionals use AI systems to make decisions on behalf of others. 
These studies mainly focus on model performance, outcome quality, and issues of fairness and accountability.
Recently, researchers have begun to examine AI-supported decision-making initiated by individuals. However, much of this research relies on \pzh{low-stake} tasks defined by the researchers. Common examples include choosing an exercise plan \cite{salimzadeh2024dealing} or planning a trip \cite{buccinca2025contrastive}.
These scenarios are often used to study trust, framing, or interpretation, but they often lack emotional depth or lasting impact. A relevant example is research on Quark Gaokao \pzh{\footnote{\url{https://vt.quark.cn/blm/pc-gaokao-1089/index?uc_param_str=dnntnwvepffrgibijbprsvdsdicheiniutstkp&x_render_type=csr&ch=pcquark@homepage_gaokao_seoredirect&entry=p_redirect}}}, a widely used AI tool for college admissions planning in China. Despite the importance of the decisions, research has found that parents often dominate the process of choosing universities, and students, while directly affected, have limited involvement \cite{10.1145/3706598.3713341}. 
\pzh{
Furthermore, previous AI-powered decision-making support tools usually rely on ground-truth data or models trained on it. 
}

In contrast, LLM fortunetelling does not rely on objective data sources for decision recommendations. Instead, it involves self-directed decisions that carry emotional and social significance. Rather than completing researcher-designed tasks, users engage with LLMs to explore uncertainty, seek guidance, and construct personal meaning. Our work contributes to understanding whether and how users’ decisions and behaviors are affected in the context of LLM fortunetelling.


\begin{figure*}
  \includegraphics[width=0.95             
 \textwidth]{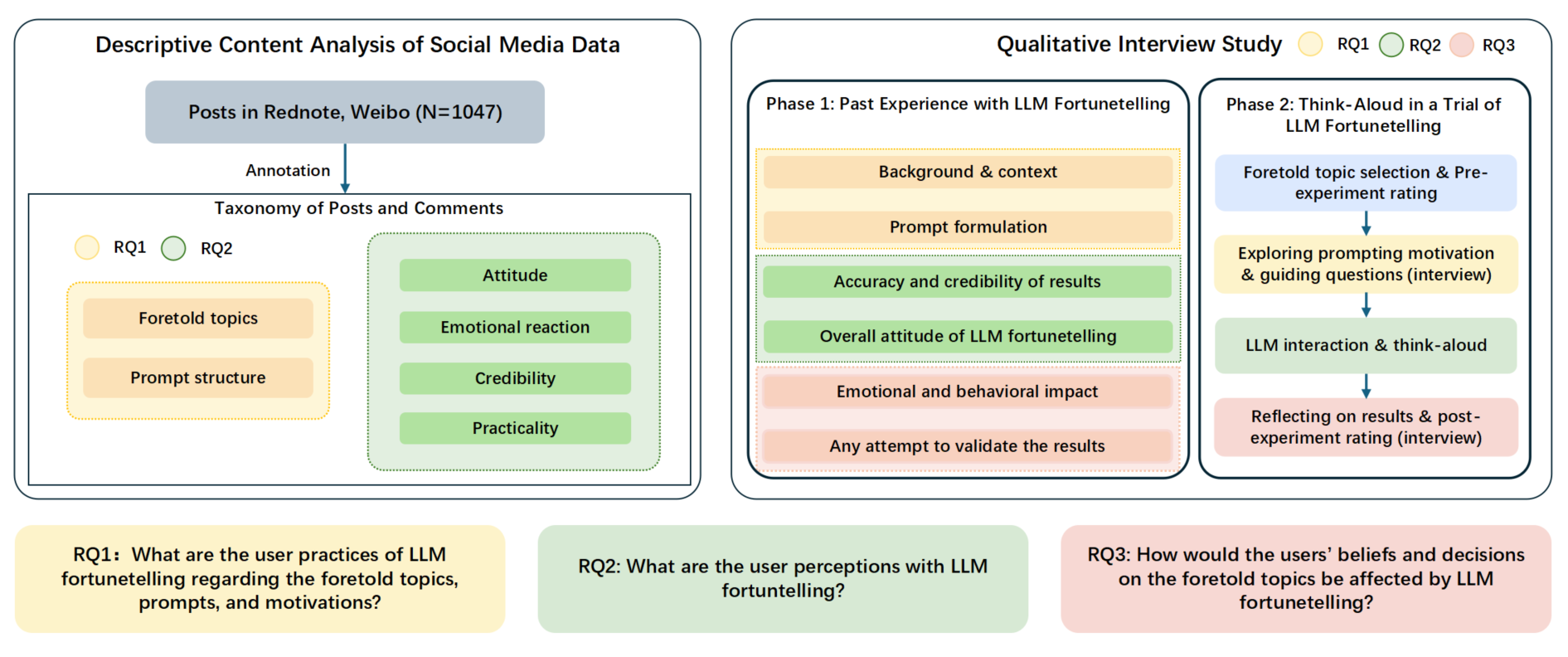}
  \caption{
  Research flow of understanding fortunetelling with LLMs by social media data analysis and interview study.}
  \label{fig:research_flow}
\end{figure*}

\section{\revise{Descriptive Content Analysis of Social Media Data about Fortunetelling with LLMs}}
\pzh{ 
We used \revise{two qualitative} studies to understand fortunetelling with LLMs.  
\penguin{\autoref{fig:research_flow} shows our research flow.}
This section presents \revise{descriptive content analysis} of posts in social media to gain a comprehensive understanding, while the next section presents an interview study to gain in-depth qualitative insights. 
Previous research has leveraged social media data to understand users' practices and opinions regarding LLM usage in specific domains like education \cite{tlili2023if}. 
Besides, prior ICWSM work has examined how social media discourse reflects and shapes user behavior and social phenomena, including issues such as online incivility and youth well-being \cite{kim2024assessing}. This line of research underscores social media as a context in which individual experiences and expressions can become publicly visible and socially consequential.
Following that, we collected data, conducted open coding, and trained classifiers, producing a comprehensive taxonomy of users' practices (RQ1) and perceptions (RQ2) of LLM fortunetelling. 
}



\subsection{Data Collection}
\pzh{
We collected data from Rednote (or Xiaohongshu) and Weibo, two major Chinese social media platforms. 
\revise{We selected these platforms because they form a complementary and widely adopted ecosystem, enabling us to capture diverse user practices surrounding LLM fortunetelling while accounting for platform-level differences through stratified analysis (see \autoref{tab:rq1_taxonomy} and \autoref{tab:rq2_taxonomy}).}
As of the second quarter of 2025, Rednote hosts about 300 million monthly active users and over 100 million contributors \footnote{\url{https://www.xiaohongshu.com/}}, primarily young users sharing lifestyle content.
In comparison, Weibo has 588 million monthly active users \footnote{\url{https://passport.weibo.com/}} and mainly supports public discussion, trending topics, and rapid information sharing.
To search posts relevant to fortunetelling with LLMs, we used combined keywords about mainstream LLMs (\eg ``DeepSeek'', ``GPT'', ``Gemini'', ``Grok'') and fortunetelling (\eg ``fortunetelling'', ``metaphysics'', ``decision'', ``choice''). }
\lxe{From Rednote, we used the Spider\_XHS crawler \footnote{\url{https://github.com/cv-cat/Spider_XHS}}
 to retrieve the top 200 posts for each combined search keyword.}
After removing duplicated posts, we initially had 1,157 posts from Rednote. 
From Weibo, we manually collected 230 text-image posts using the combined keywords in the Weibo search engine. 
\revise{To ensure comparability across platforms, we applied the same inclusion and exclusion criteria to both datasets.}
We manually went through all collected posts from Rednote and Weibo and removed 340 of them, which were irrelevant to LLM fortunetelling. 
\lxe{The removal criteria included advertisements, promotional content, and posts unrelated to fortunetelling practices with LLMs.}
Lastly, we had 1,045 posts, with 37 of them containing videos and 1,008 posts having images.
\revise{The final dataset includes 859 posts from Rednote and 186 posts from Weibo, and all subsequent analyses report both aggregated results and platform-specific distributions.}
\lxe{These posts averaged 620 likes and 66 comments per post.}
\pzh{
All data were publicly available at the time of collection. 
Following ethical research practices, we excluded private or non-public information, anonymized user identifiers, and focused exclusively on aggregated analyses to protect user privacy and minimize potential risks.}

\subsection{\pzh{Coding the Posts}}
\pzh{
We adopted an open coding approach to understand LLM-based fortunetelling shared in the posts, following established procedures in related research \cite{10.1145/3544548.3581476}. 
A post could be assigned one code in each category and multiple codes within a category (\eg prompt structure).
If a post did not belong to a code in a category, we mark it as None in that category. 
The collected posts contain videos or images, which makes it difficult to develop classifiers for automatic analyses. 
Therefore, we manually coded all 1,045 posts. 
Specifically, two authors annotated 266 posts
to develop a codebook. 
They first independently reviewed the post content and assigned codes related to our research questions, resulting in two independent codebooks. 
The two authors then collaboratively produced a final codebook through three rounds of discussions and iterations.
The two coders then re-coded the 266 posts based on the final codebook.
Throughout the coding process, we consulted relevant literature  \peng{\cite{10.1145/3544548.3581476}} and psychology research \peng{\cite{huang2024apathetic}} to guide the definition of each code. 
For example, ``Credibility'' corresponds to the commonly used practicality evaluation dimensions in AI-assisted decision-making research \cite{10.1145/3613904.3642671}.
We also identified a small number of posts discussing how LLM fortunetelling affected users’ decisions (RQ3).
Therefore, the final codebook focused on the themes of user practices (RQ1, \autoref{tab:rq1_taxonomy}) and perceptions (RQ2, \autoref{tab:rq2_taxonomy}), with an additional theme about the post topics. 
To assess the reliability of the open coding process, we calculated inter-coder agreement for each code using Cohen's Kappa, resulting in an overall mean of 0.845 (SD = 0.095). Specifically, within the ``Foretold Topics'' category (\autoref{tab:rq1_taxonomy}), the codes of ``Lost Property'' and ``Fertility'' have the highest agreement of $\kappa=1$, while within the ``Emotional Reaction'' category (\autoref{tab:rq2_taxonomy}), the ''Humorous Commentary'' code has the lowest agreement of  $\kappa=0.712$. 
After establishing the taxonomy, the two coders worked together to annotate the remaining posts. 
\revise{Finally, we obtained a labeled dataset for analysis. In the following analysis, we primarily analyzed the merged dataset from the two platforms to capture overall patterns.
As a preview of these comparisons (see \autoref{tab:rq1_taxonomy} and \autoref{tab:rq2_taxonomy}), the distributions on Rednote and Weibo are generally comparable: although some differences exist (\eg the ``Career'' topic: 24.10\% vs. 16.67\%), key categories appear on both platforms (\eg the ``General Fortune'' category: 16.18\% vs. 15.59\%), and the overall attitude trends are similar (\eg the ``Positive'' category: 58.32\% vs. 72.04\%). This supports our use of the merged dataset for analysis while preserving platform-specific nuances.
}}

\pzh{
}

\pzh{
\subsection{Theme: Topics of Posts}
Posts share three codes (\autoref{fig:three_figs}) about their topics, \ie discussion on fortunetelling (proportion for posts: 18.66\%), sharing experiences with fortunetelling (74.83\%), and sharing prompts (23.35\%). 
Within the category of discussions on fortunetelling, users’ conversations further focused on specific aspects, including reliability and safety concerns (44.78\% of discussion posts,  \eg worries that AI simply tells people what they want to hear), model performance comparisons (30.43\%, \eg debating whether Gemini or GPT provides more accurate predictions), suggestions for using LLM fortunetelling (8.70\%, \eg encouraging others to use AI to avoid bad outcomes), suggestions for prompting LLMs (10.87\%, \eg proposing to generate a birth chart first before asking the model), and other topics (5.22\%, \eg reflections on how tools like DeepSeek affect people’s sense of destiny).
\pzh{When sharing fortunetelling experiences, posters share screenshots of their fortunetelling with LLM and describe the results generated by LLM and their feelings in the post content.}
\lxe{And when sharing prompts, posters share their complete prompts for fortunetelling using LLM in the post content.}

}

\subsection{Theme: User Practices (RQ1)}
\label{sec:social_practice}
As shown in \autoref{tab:rq1_taxonomy}, we identify multiple codes that fall into the following two categories regarding user practices with LLM fortunetelling. 

\newcommand{\graycline}{%
  \arrayrulecolor{gray!60}\cline{2-5}\arrayrulecolor{black}%
}

%

\normalsize

\textbf{Foretold topics.}
\pzh{
This category refers to the topics that users would like to foretell using LLMs.
The most common topics include romance (27.08\% of posts), career (22.78\%), and general fortune (16.08\%). 
Romance-related queries often concern relationships and future partners, while career-related queries focus on job changes and career development.
General fortune-telling typically involves users providing personal information such as birth date and birthplace for holistic predictions about life, marriage, career, and wealth.
The topic of academic studies (10.91\%) is also notable. Users often consult LLMs about exam results or higher education opportunities. Wealth-related queries (9.86\%) mainly focus not on financial planning, but on luck and prosperity. 
For example, users ask when they might become wealthy or whether their ``fortune in wealth'' is favorable. 
Some users even use LLM fortunetelling to locate lost objects (3.54\%), showing its role in both major life concerns and everyday practical issues.
These findings suggest that LLM fortunetelling combines traditional concerns such as romance and wealth with modern uncertainties related to education and career, functioning as a general resource for navigating everyday uncertainty.
}

\begin{table*}[t!]
\centering
\footnotesize
\setlength{\tabcolsep}{4pt}
\caption{\revise{Distribution of foretold topics in LLM fortune-telling, identified through a content analysis of 1,045 social media posts (859 from Xiaohongshu and 186 from Weibo), with the specific quantities and frequencies for each platform.}}
\label{tab:rq1_taxonomy}
\begin{tabular}{>{\centering\arraybackslash}p{\dimexpr0.10\textwidth\relax}
                >{\centering\arraybackslash}p{\dimexpr0.12\textwidth\relax}
                p{\dimexpr0.40\textwidth\relax}
                >{\centering\arraybackslash}p{\dimexpr0.10\textwidth\relax}
                >{\centering\arraybackslash}p{\dimexpr0.10\textwidth\relax}
                >{\centering\arraybackslash}p{\dimexpr0.10\textwidth\relax}}
\toprule
\multicolumn{1}{c}{\textbf{Categories}} & 
\multicolumn{1}{c}{\textbf{Codes}} & 
\multicolumn{1}{c}{\textbf{Sample}} & 
\multicolumn{1}{c}{\textbf{Total}} & 
\multicolumn{1}{c}{\textbf{Xiaohongshu}} & 
\multicolumn{1}{c}{\textbf{Weibo}} \\
\midrule

\multirow{11}{*}{\parbox[c]{1.6cm}{\centering Foretold Topics}} 
& Career & ``The readings I did for myself and my friends about our current work situations were quite accurate, and even the predictions about career development were somewhat relevant.'' & 238 (22.78\%) & 207 (24.10\%) & 31 (16.67\%) \\
\cline{2-6}
& Romance & ``It says I am skeptical about marriage and will marry late or not marry at all.'' & 283 (27.08\%) & 256 (29.80\%) & 27 (14.52\%) \\
\cline{2-6}
& General Fortune & ``It even got my Bazi wrong.'' & 168 (16.08\%) & 139 (16.18\%) & 29 (15.59\%) \\
\cline{2-6}
& Health & ``It's actually spot on, I can't believe it even knew my right leg keeps getting injured!'' & 48 (4.59\%) & 44 (5.12\%) & 4 (2.15\%) \\
\cline{2-6}
& Academic Studies & ``Chatgpt calculated that I would be admitted and would likely receive a scholarship. Turns out I was the only one in my major who received a scholarship.'' & 114 (10.91\%) & 108 (12.57\%) & 6 (3.23\%) \\
\cline{2-6}
& Fertility & ``It was right when he said I'd have a daughter.'' & 25 (2.39\%) & 25 (2.91\%) & 0 (0.00\%) \\
\cline{2-6}
& Wealth & ``It said I'll make 150 million in middle age, I don't even dare to believe it!'' & 103 (9.86\%) & 86 (10.01\%) & 17 (9.14\%) \\
\cline{2-6}
& Interpersonal Relationships & ``The result from the AI says I need to watch out for bad people this year.'' & 38 (3.64\%) & 33 (3.84\%) & 5 (2.69\%) \\
\cline{2-6}
& Lost Property & ``I finally found the ring! DeepSeek, I'll stand by you, it said near the water source and I ended up finding it in the kitchen sink.'' & 37 (3.54\%) & 36 (4.19\%) & 1 (0.54\%) \\
\cline{2-6}
& Other Topic & ``I used DeepSeek for fortune-telling to choose which blind box to buy, and it actually picked the one I wanted!'' & 91 (8.71\%) & 85 (9.90\%) & 6 (3.23\%) \\
\cline{2-6}
& \textcolor{gray}{None} & \textcolor{gray}{-} & \textcolor{gray}{204 (19.52\%)} & \textcolor{gray}{121 (14.09\%)} & \textcolor{gray}{83 (44.62\%) }\\
\bottomrule
\end{tabular}
\end{table*}

\begin{table*}[t!]
\centering
\footnotesize
\setlength{\tabcolsep}{4pt}
\caption{\revise{This table systematically categorizes user perceptions of LLM-generated fortunetelling results into five key dimensions: Attitude, Emotional Reaction, and Assessments of Credibility and Practicality. It provides both qualitative examples and descriptive statistics (from 1,045 posts, 859 Xiaohongshu, 186 Weibo), specifically including platform-specific quantities and frequencies to offer a comprehensive overview of the user perceptions of the fortunetelling results given by LLMs (RQ2).}} 
\label{tab:rq2_taxonomy}
\begin{tabular}{
  >{\centering\arraybackslash}p{0.10\textwidth}
  >{\centering\arraybackslash}p{0.10\textwidth}
  p{0.42\textwidth}
  >{\centering\arraybackslash}p{0.10\textwidth}
  >{\centering\arraybackslash}p{0.10\textwidth}
  >{\centering\arraybackslash}p{0.10\textwidth}
}
\toprule
\multicolumn{1}{c}{\textbf{Categories}} & 
\multicolumn{1}{c}{\textbf{Codes}} & 
\multicolumn{1}{c}{\textbf{Sample}} & 
\multicolumn{1}{c}{\textbf{Total}} &
\multicolumn{1}{c}{\textbf{Xiaohongshu}} &
\multicolumn{1}{c}{\textbf{Weibo}} \\
\midrule

\multirow{4}{*}{\makecell{Attitude}} 
& Positive & ``The AI's too good at this, feels like fortune tellers are about to lose their jobs.'' 
& 635 (60.77\%) & 501 (58.32\%) & 134 (72.04\%) \\
\cline{2-6}
& Neutral & ``Some of the AI's fortunetelling is spot on, but some parts are way off.'' 
& 176 (16.84\%) & 130 (15.13\%) & 46 (24.73\%) \\
\cline{2-6}
& Negative & ``Not sure who's hyping this up, I checked it out and the fortunetelling was way off.'' 
& 66 (6.32\%) & 60 (6.98\%) & 6 (3.23\%) \\
\cline{2-6}
& \textcolor{gray}{None} & \textcolor{gray}{``DeepSeek says I'll meet my true love this year.''} 
& \textcolor{gray}{167 (15.98\%)} & \textcolor{gray}{167 (19.44\%)} & \textcolor{gray}{0 (0.00\%)} \\
\midrule

\multirow{7}{*}{\makecell{Emotional\\Reaction}} 
& Surprise & ``I asked DeepSeek what place I'd get in the interview, and it said second — and guess what, I really came in second!'' 
& 360 (34.45\%) & 310 (36.09\%) & 50 (26.88\%) \\
\cline{2-6}
& Worry & ``AI predicted that my husband and I are destined to get divorced. What should I do?'' 
& 38 (3.64\%) & 33 (3.84\%) & 5 (2.69\%) \\
\cline{2-6}
& Humorous Commentary & ``The prediction says I'll meet my true love this year—great! Can you also calculate his phone number for me?'' 
& 151 (14.45\%) & 128 (14.90\%) & 23 (12.37\%) \\
\cline{2-6}
& Hope and Encouragement & ``The prediction says that he will definitely reach out to me between 1 and 2 p.m. this afternoon—I hope it comes true!'' 
& 106 (10.14\%) & 65 (7.57\%) & 41 (22.04\%) \\
\cline{2-6}
& Contradiction & ``The AI says I'll have four kids and the first will be a girl… but my first is a boy, and I'm not having any more!'' 
& 53 (5.07\%) & 45 (5.24\%) & 8 (4.30\%) \\
\cline{2-6}
& Other Reactions & ``I'll just wait and see whether this result turns out to be accurate or not.'' 
& 16 (1.53\%) & 16 (1.86\%) & 0 (0.00\%) \\
\cline{2-6}
& \textcolor{gray}{None} & \textcolor{gray}{-} 
& \textcolor{gray}{322 (30.82\%)} & \textcolor{gray}{262 (30.50\%)} & \textcolor{gray}{60 (32.26\%)} \\
\midrule

\multirow{4}{*}{\makecell{Credibility}} 
& Skeptical & ``Why do I feel like it's saying nice things on purpose?'' 
& 377 (36.08\%) & 313 (36.44\%) & 64 (34.41\%) \\
\cline{2-6}
& Trust & ``It's really accurate, he said I'd get an offer this month and I started my job today.'' 
& 297 (28.42\%) & 264 (30.73\%) & 33 (17.74\%) \\
\cline{2-6}
& Distrust & ``It can't even get the Bazi right.'' 
& 65 (6.22\%) & 58 (6.75\%) & 7 (3.76\%) \\
\cline{2-6}
& \textcolor{gray}{None} & \textcolor{gray}{-} 
& \textcolor{gray}{303 (29.00\%)} & \textcolor{gray}{222 (25.84\%)} & \textcolor{gray}{81 (43.55\%)} \\
\midrule

\multirow{3}{*}{\makecell{Practicality}} 
& Practical & ``Last time I lost something, I told it the rules for the calculation and it helped me find it, that's really awesome.'' 
& 525 (50.24\%) & 462 (53.78\%) & 63 (33.87\%) \\
\cline{2-6}
& Impractical & ``Not accurate, it said I could score 638 but I only got 566.'' 
& 72 (6.89\%) & 66 (7.68\%) & 6 (3.23\%) \\
\cline{2-6}
& \textcolor{gray}{None} & \textcolor{gray}{-} 
& \textcolor{gray}{443 (42.39\%)} & \textcolor{gray}{328 (38.18\%)} & \textcolor{gray}{115 (61.83\%)} \\

\bottomrule
\end{tabular}
\end{table*}

\normalsize

\textbf{Prompt structure.}
\pzh{This category refers to the prompt structure used by users of LLM fortunetelling, including common character settings, output format, and so on \cite{google_prompt_generative_AI_2025}.
As shown in \autoref{tab:rq1_prompt_structure} in the Appendix, the prompt structure shared by users on social media is relatively fixed, 
combining strategies of role-playing LLMs with information needed in traditional fortunetelling practices.
Most prompts involve contextual input, such as providing birth information and astrological information, \revise{often corresponding to specific traditional systems such as Bazi and Ziwei Doushu, as well as Western astrology.}
\revise{Our dataset shows that Chinese traditional methods dominate these practices, with Bazi/Four Pillars appearing in 215 posts (20.57\% of all posts; 65.95\% among posts that explicitly mention a method), followed by Ziwei Doushu (3.45\%) and other systems such as Liuyao (3.25\%). In contrast, Western methods such as astrology (5.55\%) and tarot (2.20\%) appear less frequently.}
Meanwhile, most prompts provide complete task instructions, such as \textit{``Analyze my horoscope.''}
Users also frequently use character settings, instructing the LLM to act as a professional fortune-teller, and knowledge recall (15.22\%), citing classic texts or traditional methods. 
These practices place interactions with LLM within a familiar cultural framework, indicating that users view LLM as a continuation of traditional fortunetelling systems.}

\pzh{
\subsection{Theme: User Perceptions (RQ2)}
\label{sec:social_perceptions}
\autoref{tab:rq2_taxonomy} summarizes the user perceptions of LLM fortunetelling shared in the posts, which fall into the following categories.

\textbf{General attitude}. 
\pzh{This category refers to the user's overall attitude towards LLM fortunetelling, including positive, neutral, and negative attitudes.
Most users (60.77\% of posts) hold a positive attitude toward LLM fortunetelling in social media, 
often praising the accuracy of predictions or joking about LLM replacing human fortune-tellers. 
Neutral opinions are also common (16.84\%), while explicitly negative views remain in the minority (6.32\%), mainly criticizing inaccurate results or the LLM’s tendency to cater to users’ expectations. 
Overall, users generally engage with LLM fortunetelling positively while remaining aware of its limitations.}

\textbf{Emotional reactions}. 
\lxe{This category refers to the user's emotional reaction (\eg surprise, worry) to the LLM fortunetelling results as reflected in the post.
The most emotional reaction of users (34.45\% of posts) to fortunetelling results is surprising when the prediction matches reality.
Many users regard this coincidence as validation, thereby increasing their investment in the results.
Additionally, some users interpret predictions humorously (14.45\%). For example, they joke about the irrationality of the results, using a relaxed tone to deflect their seriousness. 
Other users (10.14\%) use predictions as a source of psychological support, offering comfort or encouragement and hoping for a positive outcome. 
Meanwhile, a small number of users perceive contradictions, such as concerns about unfavorable marriage or career prospects (3.64\%), demonstrating that LLM fortunetelling can sometimes trigger negative emotions.
Overall, these emotional reactions reflect both users’ expectations and entertaining interpretations of the results, while also revealing their psychological impact.}

\textbf{Perceived credibility}. 
\lxe{This category captures how users assess the credibility of LLM fortunetelling results, ranging from trust to skepticism and outright distrust. Among posts, 36.08\% convey rational skepticism, often asking if the system is saying nice things on purpose or deliberately catering to user expectations. Meanwhile, 28.42\% of posts express a strong trust, with users reporting that predictions aligned with real-life experiences, such as receiving a job offer after being told their career luck is improving. A smaller proportion (6.22\%) voices a clear distrust, often pointing out obvious inaccuracies, for example, complaining that the system could not even get the Bazi right. Overall, these findings suggest that while many users view LLM predictions as somewhat credible, a notable portion remains doubtful or critical.}

\textbf{Perceived practicality}. 
\lxe{Perceived practicality concerns whether users regard LLM fortunetelling as useful in everyday life. About half of the posts (50.24\%) describe it as practical, noting applications such as finding lost items or guiding routine decisions, while 6.89\% of posts criticize its lack of accuracy and question its real-world value. At the same time, 42.39\% of posts do not explicitly address usefulness, reflecting that systematic reflections on utility are less frequent than personal anecdotes. Overall, users tend to perceive LLM fortunetelling as practical when it provides help in specific situations, but they rarely engage in broader or systematic evaluations of its overall value.}
}

\pzh{
\section{Qualitative Interview Study with Users of LLM Fortunetelling}
While the \revise{descriptive content analysis} of social media data provided a comprehensive taxonomy of users' practices (RQ1) and perceptions (RQ2) of LLM fortunetelling, we still lack in-depth insights into their underlying reasons and the impact on users' decision-making behaviors (RQ3).  
Therefore, we further conducted semi-structured interviews with 20 users of LLM-based fortunetelling, focusing on how they interpret and respond to the fortunetelling results. \cite{liu2025understand}
}
\subsection{Participants}
\pzh{
We recruited 7 creators of our collected Rednote posts who accepted our invitation sent via private messages (we sent \peng{67} invitations in total). 
We also used two principles to select 13 out of 100 respondents to our online advertisements in the authors' social networks (\eg via WeChat group chats). \peng{First, they should have experience in LLM fortunetelling. 
Second, their occupations or majors and foretold topics in the past should be diverse.}
In the invitations or advertisements, individuals would receive a detailed overview of the study, including its purpose, what participation involves, the expected duration of their involvement, any potential risks and benefits, and their rights as participants. After confirming the consent, participants would engage in a one-hour semi-structured online interview. Each participant received a cash compensation of 50 Chinese Yuan (equivalent to $6.92$ USD) after finishing the interview.
\autoref{tab:participants} in the Appendix describes the detailed information of our 20 participants (P1-20, 15 females and 5 males, age range: 20-43). 

}

\pzh{
\subsection{Procedure}
The interview study consisted of two phases: one asks about the participants' past experiences with LLM fortunetelling; the other seeks to learn their in-situ thoughts in a trial of LLM fortunetelling.  
\autoref{tab:combined} in the Appendix summarizes the questions we asked in two phases, respectively. 

\subsubsection{Phase 1: Past Experience with LLM Fortunetelling}
To get started, participants are asked to recall their previous experiences using LLMs for fortunetelling. 
We ask their foretold topics, motivations, prompting strategies (RQ1), perceptions of the results (RQ2), and their subsequent thoughts and behaviors (RQ3). 
Key questions include the context in which participants used LLM fortunetelling, their prompt formulation, immediate reactions to the LLM responses, perceived accuracy and credibility, and overall attitudes. 
Specifically, we have questions asking about the emotional and behavioral impact of the fortunetelling results on participants, as well as any attempt to validate the results. 

\subsubsection{Phase 2: Think-Aloud in a Trial of LLM Fortunetelling}
Following phase 1, participants select a foretold topic (\eg romance, academic studies) that is closely related to them at that moment to conduct LLM (mostly DeepSeek) fortunetelling with LLMs on their computers. 
We prepare example prompts sourced from the related posts on social media platforms for reference. 
This trial can provide a fresh experience of LLM fortunetelling to understand how participants interpret and are affected by the results. 
Specifically, before and after the trial, we ask participants to rate their likelihood of making a specific decision (\eg quit a job or not) or level of confidence towards the foretold topics in the future (\eg paper acceptance) from 0 to 100\%. 
Prior to the trial, we ask participants about their prompting motivation and present them with guiding questions asking about their immediate emotional reactions to the LLM answers (\eg feeling understood, encouraged, or confused), perceived strengths or shortcomings of the answers, and potential behavioral impact. 
During the trial, participants are asked to think aloud, expressing their thoughts related to the questions above in real time. 
All interactions are screen- and audio-recorded, anonymized, and treated confidentially. 
For the interview recordings, two of the authors transcribed them into text and conducted a thematic analysis. They first familiarized themselves by reviewing all the text scripts independently.
After several rounds of coding with comparison and discussion, they finalized the codes of all the interview data. 
We counted the occurrences of codes and incorporated these qualitative findings in the following presentation of our results.
}

\pzh{
\subsection{Findings}
}

\pzh{
\subsubsection{Motivations (RQ1): decision support, emotional comfort, entertainment, and social substitution.}
Users turned to LLM fortune-tellers when making critical decisions, such as furthering their studies (P7) or resignation (P9), seeking guidance and reassurance. 
Meanwhile, participants frequently emphasized the dimension of emotional comfort by using LLM fortunetelling, \eg to seek support before exams or competitions (P1, P18) or after conflicts in close relationships (P4, P6). 
 In fact, many participants mentioned that they primarily tried fortunetelling out of curiosity or entertainment. 
 Some users also used LLM fortunetelling to confirm pre-existing thoughts (P17) or as a substitute for friends or counselors (P16), fulfilling needs for communication and companionship.
 One participant (P11) even described consulting LLM fortunetelling as a habitual step before making decisions.
 }

 \pzh{
}

\pzh{
\subsubsection{Participants exhibited highly different perceptions of LLM fortunetelling results, which are often presented in a partially truthful narrative style (RQ2).} 
In the experiment, some users found the results reliable, particularly convincing when using horoscopes to accurately predict the timing of reconciliation with a boyfriend (P10) or academic milestones (P13). 
A few results closely matched their personal experiences, \eg the family experiencing a flood (P5) and exchanging programs in Hong Kong (P8), leaving users shocked. 
However, many users (P2, P3, P7, P10, P14, P17, P18) found the results vague and general without specific operational guidance, making them still feel uncertain about their decisions. 
 During the trials of LLM fortunetelling during the study, users often found themselves with a sense of partial conviction and partial skepticism. 
 They were surprised by the corrected information in LLM's responses, but they also identified obvious errors or ambiguous language. \textit{``When I noticed that the AI miscalculated my age, I felt it was foolish.'' (\peng{P18})} 

 }

\pzh{
\subsubsection{Participants have neutral or positive attitudes towards LLM fortunetelling but are aware of the LLM's tendency to cater to their expectations (RQ2).} 
\label{sec:inter_attitude}
Some participants pointed out potential concerns of LLM fortunetelling, such as over-reliance (\eg relying too heavily on LLM fortunetelling results for important decisions) and privacy issues (\eg disclosing birth data or sensitive personal experiences). 
Nevertheless, most users regarded LLM fortunetelling as harmless and primarily treated it as a form of entertainment or psychological comfort. 
In the trials during the study, seventeen users found that the LLM often tailors its responses to align with their tone and expectations. 
For example, the responses often contained compliments (\eg ``\textit{You'll achieve great success in your future career}'') or initially presented challenges but ultimately concluded positively (\eg ``\textit{You may face some setbacks right now, but things will soon turn around}''). 
Interestingly, P9 and P19 said that they expected or readily accepted LLM's catering in fortunetelling, as it fostered confidence and hope, alleviating anxiety and uncertainty in the short term. 
Conversely, \peng{P4 and P7} preferred the LLM to maintain a neutral, rational, or objective tone to give fortunetelling results, saying that this can enhance the results' credibility and reference value.
}

\subsubsection{\revise{LLM fortunetelling exerts modest influences on users’ beliefs and decision-making processes (RQ3).}}
\label{sec:rq3_finding_1}
Our findings suggest that LLM fortunetelling mainly provides psychological reassurance and modestly influences the timing of decisions rather than overall directions \cite{ma2024you}.

\textbf{Influence on beliefs and emotions.}  
Several participants (P1, P10, P11, P12, P19) described using LLM fortunetelling as a form of psychological adjustment. It offers reassurance in times of stress or anxiety, and provides narratives that resonate with their inner struggles. 
For instance, P11 consulted the LLM when feeling miserable at work, and upon receiving a reframe—that the feeling stemmed from a longing for freedom constrained by reality, and that work could be approached as an ``observation game''—found the perspective genuinely helpful and reported feeling less drained by office politics thereafter.
Additionally, in the experiment, 17 participants chose a topic to predict and rated their confidence in predictive outcomes generated by LLMs, covering both broad life domains (\eg romance, career) and specific personal goals (\eg whether she could successfully lose weight). As shown in the \autoref{tab:participants} in the Appendix, five participants showed no change before and after the interaction, while the remaining exhibited an average change of 9.53 points (SD = 8.84), with the largest shift being 30 points and the smallest 0. 
\revise{These results suggest that LLM fortunetelling can calibrate the degree of certainty in participants’ beliefs about the future.}


\textbf{Influence on decisions.}
By contrast, the direct influence of LLM fortunetelling on concrete decision-making is limited. While most participants emphasized that they relied primarily on personal judgment, financial conditions, and family considerations, the advice generated by LLM sometimes shaped the timing of actions. For example, participants reported delaying resignation (P9), postponing competitions (P18), or suspending investments (P16) based on LLM suggestions. 
P9, who was already strongly considering leaving their job, consulted the LLM about career prospects and upon being advised not to quit in 2025—an output that matched existing worries—chose to hold on.
\lxe{
Rather than changing decision directions, LLM fortunetelling mainly reinforced participants' existing inclinations and influenced the timing of actions. Participants often felt reassured when the generated results aligned with their preferred choices.
For example, P17 entered the consultation already leaning toward staying in the current department, and the fortunetelling result unexpectedly aligned with this inclination, reinforcing the sense that remaining might be the more reasonable choice.} 
This shows that LLM fortunetelling did not necessarily change participants’ decision directions but strengthened their beliefs in the legitimacy of their own preferences.
\revise{Importantly, these effects do not indicate changes in decision outcomes, but rather influence the timing and pacing of actions.}

\subsubsection{Users actively interpret, filter, and internalize LLM-generated results, and these practices influence how fortunetelling shapes their beliefs about everyday life (RQ3).} 
Rather than passively accepting results, participants reinterpret vague predictions as actionable or psychologically useful guidance. For example, P12 transformed ``pay attention to your kidneys” into “stay up less late''. At the same time, P11 summarized general health warnings as reminders to monitor their diet. Some even integrated generic outputs into specific plans, such as aligning career (P9) or exam schedules (P19) with the predicted time. 
\lxe{The impact is therefore selective. Short-term and concrete predictions can foster small belief adjustments and motivate behavioral fine-tuning. In contrast, long-term or ambiguous predictions are approached with skepticism and mainly provide psychological reassurance.}

\section{Discussion}
\pzh{In this work, we offer comprehensive findings on user practices, perceptions, and impacts of LLM fortunetelling via analyzing social media data and interviewing users.
This section discusses the implications of these findings for related research fields, values and risks of LLM fortunetelling, as well as limitations and future work.  
}

\revise{
\lxeee{\subsection{Theoretical Explanation of Findings}}
\revise{Our findings on LLM fortunetelling can be theoretically explained through the lenses of motivated reasoning, confirmation bias, and the Barnum effect.
Users approached LLM fortunetelling driven by desires for reassurance, validation, and emotional comfort. This motivational basis reflects motivated reasoning \cite{kunda1990case}, which posits that individuals tend to interpret information in a way that aligns with their own desires and emotional needs—this explains why users consistently seek outcomes that conform to rather than contradict their inner inclinations in various matters, from career transitions to interpersonal relationship choices.}

\revise{
The ways in which users perceived and interpreted LLM outputs further reveal confirmation bias \cite{nickerson1998confirmation} and the Barnum effect \cite{huizi2022understanding} at work. 
Participants selectively associated vague predictions with personal experiences and reinterpreted ambiguous outputs as actionable guidance.
This resonates with the Barnum effect, where generic statements feel personally meaningful precisely because individuals project their own experiences onto them.
At the same time, confirmation bias led users to favor outputs that aligned with their prior beliefs—a tendency that manifests across seeking, interpreting, and recalling information, and is especially pronounced among individuals with stronger prior beliefs \cite{10.1145/3706598.3713873}.
Notably, seventeen participants observed that the LLM tailored its responses to match their tone and expectations, yet largely accepted this catering: P9 and P19 even anticipated and welcomed it, as it fostered confidence and alleviated anxiety. 
This pattern echoes findings in human-AI collaboration, where AI outputs that mirror users' initial judgments boost users' confidence in those judgments and amplify confirmation bias \cite{rosbach2025two}.

}
Because users selectively interpreted LLM outputs to align with their existing beliefs, the influence on their actual decisions was modest but consistent.
Rather than altering decision directions, LLM fortunetelling primarily reinforced existing inclinations—P9 held off on resignation because the LLM output matched existing worries. Users adjusted the timing of actions when outputs aligned with prior preferences, and overall confidence in existing beliefs shifted by an average of 9.53 points. This pattern resembles positive illusions \cite{taylor1988illusion}, where selectively internalized information sustains a sense of control over uncertain futures.
Taken together, LLM fortunetelling functions less as a predictive tool and more as a space for emotional reflection and meaning-making \cite{10.1145/3706598.3713453}, reconfiguring traditional fortunetelling into a hybrid cultural practice that combines belief reinforcement with LLM-mediated emotional support for navigating everyday uncertainty.
}

\subsection{Platform Dynamics and Social Implications of LLM Fortunetelling}
LLM fortunetelling operates not only as an individual interaction with LLM, but also as a platform-mediated social practice shaped by social media. By sharing LLM-generated fortunes in public posts, users turn private uncertainty into visible narratives that circulate within platform environments, allowing fortunetelling experiences to be encountered and interpreted by broader audiences.
Social media platforms amplify certain fortunetelling narratives, particularly those that are emotionally resonant or perceived as coincidentally ``accurate.'' Features such as algorithmic feeds \cite{Jose_Geeng_Morales_McCoy_Greenstadt_2025,mousavi2024auditing} increase the visibility of these cases, 
\revise{which may contribute to a skewed impression of the reliability of LLM-generated fortunes,}
even when individual users remain skeptical. 
\revise{Such visibility may contribute to the perception that LLM fortunetelling is a socially acceptable way of seeking reassurance under uncertainty.}

\revise{The public framing of fortunetelling may also blur the boundary between entertainment and meaningful guidance.}
The coexistence of joking and serious use within the same social media context allows users to oscillate between detachment and belief, while experiencing the accumulated emotional and cognitive impact.
\revise{From a platform perspective, LLM fortunetelling is associated with how uncertainty is discussed and shared in online contexts, pointing to potential implications beyond individual interactions.}
\pzh{
}

\pzh{
}

\pzh{
}

\subsection{Ambiguity, Ingratiation, and Risk in LLM-Based Fortunetelling}
Our findings suggest that users often underestimate the risks associated with LLM-based fortunetelling. 
While some participants expressed concerns about dependence or privacy breaches, many overlooked the longer-term consequences of misleading outputs, repeated reliance, and the disclosure of sensitive personal information.
This echoes prior work showing that users frequently struggle to recognize algorithmic harms in everyday interactions \cite{10.1145/3706598.3713199}.
\revise{One factor that may contribute to these risks is the ambiguous and ingratiating nature of LLM-generated fortunetelling outputs.}
Consistent with previous research \cite{han2025tracing}, these responses tend to rely on vague but affirmative language to create a feeling of being understood or accurately ``predicted,'' even without factual grounding. 
\revise{While such responses may provide emotional reassurance during uncertainty, they may also reduce critical scrutiny and encourage overestimation of reliability, particularly when users repeatedly reference divination results.}

Current mitigation strategies, such as brief disclaimers (\eg ``for reference only''), may have limited effectiveness.
Prior work \cite{10.1145/3706598.3714097} suggests that mitigating over-reliance requires supporting users’ reflective engagement with LLM-generated guidance.
\revise{Specifically, these interventions might include embedded reflective prompts (\eg ``What other factors are important in your decision-making?'') and structured journaling features to invite users to record their thoughts along with the prediction.
Additionally, contextual framing cues (\eg differentiating between ``entertainment'' and ``decision support'' modes) and adaptive feedback (\eg showing how the prediction affects user confidence or emotions) could further encourage critical reflection.
We also suggest clearer warnings and privacy prompts when users share sensitive information.
Beyond interface design, technological solutions can be explored, such as fine-tuning models or integrating prompt-based detection mechanisms to automatically identify sensitive or privacy-related user inputs, and applying filtering or rewriting modules to sanitize such inputs before inference \cite{biswas2026guardrails}.}

Taken together, these findings highlight a paradox in LLM fortunetelling: while emotionally supportive outputs may reassure users, repeated reliance on such advice may also weaken careful scrutiny.

\pzh{
\subsection{Limitations and Future Work}
This study has several limitations. 
First, our social media data is limited to public posts \revise{on RedNote and Weibo}, which may miss private or sensitive use scenarios. 
\revise{As a result, users who engage in LLM fortunetelling without publicly sharing their experiences are not captured, potentially introducing sampling bias toward more expressive or socially motivated users. Second, although we pool data from RedNote and Weibo to identify overall patterns of LLM fortunetelling, platform-specific affordances may still shape user behavior. }
\revise{In addition, our interview sample of 20 participants may not fully represent demographic diversity and may be subject to self-selection bias, as participation was voluntary and participants were partly recruited through the authors’ networks. 
Future research could expand the sample size, recruit more diverse participants, and explore semi-private interactions to capture a broader range of experiences.}
\revise{Additionally, our social media data captures only observable behaviors, and our interviews provide a single-point, self-reported perspective (\eg short-term confidence ratings). 
As a result, we cannot assess the long-term effects of LLM fortunetelling on users. 
}

 Moreover, our analysis focuses on posts as acts of public self-expression rather than interactional dynamics in comment threads. 
\revise{In addition, our approach relies on descriptive content analysis and does not incorporate computational modeling. Future work can further conduct quantitative analyses (\eg regression) to examine the relationship between the content of LLM fortunetelling posts and the types of comments they receive. Such analyses could deepen our understanding of how different content features shape user engagement and responses.}
Finally, this study examines Chinese users and fortunetelling mostly using Chinese traditional terminology. Cultural factors may affect users' perceptions and practices of LLM fortunetelling. Future research should explore other cultural contexts to assess the broader applicability of the results.
}

\section{Conclusion}
\pzh{
We investigate how people engage in fortunetelling with LLMs through two qualitative studies, \revise{combining a descriptive content analysis of 1,045 posts on Chinese social media with interviews of 20 LLM fortunetelling users.} 
\revise{Our findings suggest that users turn to LLM fortunetelling for emotional support and decision-related reflection, and that such interactions may shape users’ beliefs and decision-making in subtle ways.}
Overall, this work positions fortunetelling as a meaningful form of daily interaction with LLMs. 
\revise{Finally, this study highlights the potential impact of LLM fortunetelling, particularly related to users' emotional experience, interpretation of the results, and perceptions of associated risks.}
}

\bibliography{main}

\section{Ethics Checklist}
\begin{enumerate}

\item For most authors...
\begin{enumerate}
    \item  Would answering this research question advance science without violating social contracts, such as violating privacy norms, perpetuating unfair profiling, exacerbating the socio-economic divide, or implying disrespect to societies or cultures?
    Yes. The social media data used in this study are anonymized and processed to protect user privacy, and no personally identifiable information is analyzed. Participants involved in interviews or surveys are treated with confidentiality and informed consent, ensuring their privacy is fully protected. The study avoids unfair profiling or cultural bias and does not contribute to social or economic inequality.
  \item Do your main claims in the abstract and introduction accurately reflect the paper's contributions and scope?
    Yes. For example, our main claims focus on the LLM fortunetelling, which accurately reflect the paper’s scope.
   \item Do you clarify how the proposed methodological approach is appropriate for the claims made? 
    Yes. We provide motivation, goal, or references for each used approach.
   \item Do you clarify what are possible artifacts in the data used, given population-specific distributions?
    Yes. The study acknowledges potential artifacts arising from population-specific distributions in the data. To mitigate this, the sample is designed to include participants from diverse age groups, occupations, and backgrounds. However, despite these efforts, some bias related to data availability and platform usage patterns may remain and cannot be completely avoided.
  \item Did you describe the limitations of your work?
    Yes.
  \item Did you discuss any potential negative societal impacts of your work?
    Yes.
      \item Did you discuss any potential misuse of your work?
    Yes. We discuss the potential misuse of LLM fortunetelling in the section ``discussion''.
    \item Did you describe steps taken to prevent or mitigate potential negative outcomes of the research, such as data and model documentation, data anonymization, responsible release, access control, and the reproducibility of findings?
    Yes. The study clearly describes multiple steps taken to prevent or mitigate potential negative outcomes. These include thorough data anonymization, careful data and model documentation, controlled access to sensitive resources, and responsible release of research outputs.
  \item Have you read the ethics review guidelines and ensured that your paper conforms to them?
    Yes.
\end{enumerate}

\item Additionally, if you are using existing assets (e.g., code, data, models) or curating/releasing new assets, \textbf{without compromising anonymity}...
\begin{enumerate}
  \item If your work uses existing assets, did you cite the creators?
    Yes. We cite them in reference or urls to the
official websites.
  \item Did you mention the license of the assets?
    No. However, we have confirmed that all the assets in our paper can be used for academic purposes.
  \item Did you include any new assets in the supplemental material or as a URL?
    No. This study does not include any new assets in the supplemental material or as external URLs. All materials used are either described within the paper or based on existing, publicly available resources, ensuring clarity and completeness without introducing additional external assets. 
  \item Did you discuss whether and how consent was obtained from people whose data you're using/curating?
    Yes. The study states that the data were collected from publicly available social media content in compliance with platform policies, and therefore individual consent was not explicitly obtained. To mitigate ethical concerns, all data were anonymized and analyzed only in aggregate for research purposes.
  \item Did you discuss whether the data you are using/curating contains personally identifiable information or offensive content?
    Yes. The paper clarifies that the data do not contain personally identifiable information, as all user identifiers are removed during preprocessing. Potentially offensive content is filtered or handled cautiously, and the analysis focuses on aggregated patterns rather than individual expressions.
\item If you are curating or releasing new datasets, did you discuss how you intend to make your datasets FAIR?
Yes. The paper discusses that the curated dataset follows FAIR principles by ensuring clear documentation and metadata for findability and reusability, standardized data formats for accessibility and interoperability, and well-defined usage conditions to support responsible reuse.
\item If you are curating or releasing new datasets, did you create a Datasheet for the Dataset? 
Yes.
\end{enumerate}

\item Additionally, if you used crowdsourcing or conducted research with human subjects, \textbf{without compromising anonymity}...
\begin{enumerate}
  \item Did you include the full text of instructions given to participants and screenshots?
    Yes. The instructions provided to participants are summarized in the paper in a way that preserves anonymity.
  \item Did you describe any potential participant risks, with mentions of Institutional Review Board (IRB) approvals?
    The local institutions of the authors do not require nor provide ethical reviews for conducting human-subjects studies like us, which are not in medical do- mains. Nevertheless, we took several ways to mitigate the ethical concerns.
  \item Did you include the estimated hourly wage paid to participants and the total amount spent on participant compensation?
    Yes. Information about participant compensation, including estimated hourly wage and total cost, is reported.
   \item Did you discuss how data is stored, shared, and deidentified?
   Yes. The study describes how data are securely stored, deidentified, and shared only under controlled conditions.
\end{enumerate}

\end{enumerate}

\section{Acknowledgments of the Use of AI}
We used AI (in particular large language models) for the following: supporting participants to experience LLM fortunetelling the during the qualitative interview study. 
Details can be found in the relevant sections. 
Authors take responsibility for the output and use of AI in this paper.

\clearpage
\onecolumn
\section{Appendix}

\begin{table}[h]
  \centering
  \caption{Interview questions in Phase 1 and Phase 2 and their corresponding research questions.}
  \label{tab:combined}
  \small
  \begin{tabular}{m{0.84\textwidth} >{\centering\arraybackslash}m{0.12\textwidth}}
    \toprule[1.5pt]
    \thead{Question Content} & \thead{Corresponding RQ} \\
    \midrule[1.2pt]

    \multicolumn{2}{c}{\textbf{Phase 1: Past Experience with LLM Fortunetelling}} \\
    \midrule

    Have you ever asked LLM to provide you with important advice (e.g., on love, academics, or career) or use LLM for fortunetelling? 
    & RQ1 \\
    \midrule

    Regarding your experience of LLM fortunetelling: & \\
    What was the background (reason) at that time? & RQ1 \\
    How was LLM's response? & RQ2 \\
    What was your intuitive feeling about the result given by LLM? Did you think it was highly credible? & RQ2 \\
    Do you think LLM fortunetelling had an impact on your subsequent mindset or thoughts? Can you give an example? & RQ3 \\
    Did these results make you rethink your choices, pause your original plans, or take new actions? & RQ3 \\
    Did you verify whether the AI fortunetelling was accurate afterward? What was the result? & RQ2 \\
    \midrule

    In what situations do you usually want to seek fortunetelling help? Is it for making decisions, emotional comfort, purely out of curiosity or entertainment? 
    & RQ1 \\
    \midrule

    When you usually use LLM for fortunetelling, how do you organize your questions (prompts)? For example, do you refer to certain traditional fortunetelling terms? 
    & RQ1 \\
    Where do you get these terms? For instance, from social media searches, or do you come up with them yourself? 
    & RQ1 \\
    \midrule

    What is your overall attitude toward LLM fortunetelling—positive, neutral, or negative? 
    & RQ2 \\
    Does it pose any harm to your life? 
    & RQ2 \\
    \midrule

    What do you think is the source of LLM fortunetelling's accuracy? Is it more like psychological suggestion, probability analysis, ``actually a bit effective,'' or something else? 
    & RQ2 \\
    \midrule

    What words would you use to describe the experience brought by LLM fortunetelling? (e.g., reassuring, fun, understood, confused, distrustful, etc.) 
    & RQ2 \\
    \midrule

    If LLM fortunetelling gives a conclusion you are unwilling to accept, how will you deal with it? Will you ignore it, adjust yourself, or seek a new interpretation? 
    & RQ3 \\
    \midrule[1.2pt]

    \multicolumn{2}{c}{\textbf{Phase 2: Think-Aloud in a Trial of LLM Fortunetelling}} \\
    \midrule

    Experiment Pre-/Post-Scoring (e.g., decision probability for ``changing jobs''; confidence scores for love/career/health) 
    & RQ3 \\
    \midrule

    What kind of answer do you hope to get from this fortunetelling? (e.g., emotional comfort, decision advice, curiosity satisfaction) 
    & RQ1 \\
    \midrule

    What's your first reaction to the answer? (e.g., shocked by accuracy, comforted, encouraged, advised, confused, etc.) 
    & RQ2 \\
    What makes you feel ``accurate'' or ``unreliable''? Why? 
    & RQ2 \\
    Do you have new questions or want to ask further? 
    & RQ2 \\
    \midrule

    Do you think these contents affect your current thoughts or decisions? 
    & RQ3 \\
    Did they change your original plans or add new options? 
    & RQ3 \\
    Do you think you will refer to these suggestions in reality? 
    & RQ3 \\
    \midrule

    Why did the score change? Can you review and explain? 
    & RQ3 \\
    \midrule

    Do you think the LLM's answer caters to you, i.e., deliberately saying nice things? 
    & RQ2 \\

    \bottomrule[1.5pt]
  \end{tabular}
\end{table}

\begin{table*}[t!]
\centering
\renewcommand{\arraystretch}{0.9}
\scriptsize
\setlength{\tabcolsep}{3pt}
\caption{Among the detailed demographics of the 20 LLM-fortunetelling interview participants, ``LLM usage exp.'' indicates how frequently they use large language models in daily life, while ``LLM fortunetelling exp.'' indicates how frequently they employ LLMs for fortunetelling. Under ``Fortunetelling Topics,'' the four columns represent different aspects: ``In the Past'' lists the topics participants mentioned when recalling previous fortunetelling experiences, ``Past Type'' indicates the type of fortunetelling in participants’ past experiences, ``During Experiment'' lists the topics they actually used in the study, and ``Rating Type'' indicates the type of fortunetelling in the experiment. ``Belief'' means confidence in a predictive outcome (\eg ``Will my paper be accepted?''), while ``Decision'' means the likelihood of taking a specific action (\eg ``Will I transfer departments?'').}
\label{tab:participants}

\rowcolors*{2}{gray!10}{white} 

\begin{tabular}{
  >{\centering\arraybackslash}m{0.4cm}   
  >{\centering\arraybackslash}m{0.6cm}   
  >{\centering\arraybackslash}m{0.4cm}   
  >{\centering\arraybackslash}m{1.85cm}  
  >{\centering\arraybackslash}m{1.0cm}   
  >{\centering\arraybackslash}m{1.4cm}   
  >{\centering\arraybackslash}m{2.2cm}   
  >{\centering\arraybackslash}m{0.9cm}   
  >{\centering\arraybackslash}m{2.2cm}   
  >{\centering\arraybackslash}m{0.8cm}   
  >{\centering\arraybackslash}m{0.4cm}   
  >{\centering\arraybackslash}m{0.4cm}   
}
\toprule
\textbf{No.} &
\textbf{Gender} &
\textbf{Age} &
\textbf{Occupation} &
\textbf{LLM usage exp.} &
\textbf{LLM fortunetelling exp.} &
\multicolumn{4}{c}{\textbf{Fortunetelling Topics}} &
\makecell{\textbf{Confidence}\\\textbf{in Decision}} \\
 & & & & & &
\textbf{In the Past} &
\textbf{Past Type} &
\textbf{During Experiment} &
\textbf{Rating Type} &
\textbf{Pre} &
\textbf{Post} \\
\midrule
1 & Female & 20 & Student in Finance & Frequently & Often & Exam prediction & Belief & Career direction & Belief & 15 & 15 \\
2 & Male   & 22 & Student in Finance & Often & Often & General fortune prediction & Decision & Romantic fortune prediction & Belief & 15 & 15 \\
3 & Male   & 25 & Student in Journalism & Frequently & Often & PhD decision & Decision & Paper acceptance prediction & Belief & 75 & 85 \\
4 & Female & 20 & Student in Sociology & Often & Occasionally & Breakup prediction & Belief & Academic and career planning & Belief & 80 & 80 \\
5 & Male   & 20 & Student in Artificial Intelligence & Often & Occasionally & Earthquake probability prediction & Belief & General fortune prediction & Belief & 70 & 80 \\
6 & Female & 23 & Student in Geomatics Engineering & Often & Often & Romantic fortune prediction & Belief & General fortune prediction & Belief & 90 & 95 \\
7 & Female & 23 & Student in Law & Frequently & Frequently & Graduate study prediction & Belief & Love timing prediction & Belief & 65 & 70 \\
8 & Female & 20 & Student in Law & Often & Frequently & Lost item divination & Belief & Academic prospect prediction & Belief & 60 & 80 \\
9 & Female & 23 & Technology company employee & Frequently & Frequently & Resignation time prediction & Belief & Promotion and salary raise prediction & Belief & 100 & 100 \\
10 & Female & 26 & Education industry employee & Frequently & Frequently & Civil service job selection & Decision & Job change decision & Decision & 45 & 50 \\
11 & Female & 23 & Technology company employee & Frequently & Frequently & Work emotion analysis & Decision & Weight loss goal prediction & Belief & 80 & 60 \\
12 & Female & 39 & Self-employed individual & Often & Occasionally & Annual fortune prediction & Belief & Earning potential prediction & Belief & 60 & 90 \\
13 & Female & 20 & Student in Law & Often & Often & General fortune prediction & Belief & Academic prospects prediction & Belief & 50 & 60 \\
14 & Male   & 20 & Student in Finance & Frequently & Often & Graduate program selection & Decision & Academic prospects prediction & Belief & 80 & 90 \\
15 & Male   & 31 & Technology company employee & Frequently & Occasionally & Civil Service exam prediction & Belief & Career achievement prediction & Belief & 50 & 60 \\
16 & Female & 43 & Self-employed & Often & Occasionally & Investment decision & Decision & Work prospect prediction & Belief & 20 & 40 \\
17 & Female & 23 & State-owned enterprise employee & Frequently & Often & Daily fortune prediction \& Travel decision & Decision & Job transition decision & Decision & 80 & 90 \\
18 & Female & 20 & Student in Applied Meteorology & Often & Frequently & Competition participation decision & Decision & Romance fortune prediction & Belief & 100 & 100 \\
19 & Female & 20 & Media sector employee & Occasionally & Occasionally & Career path decision & Decision & Career development prediction & Belief & 70 & 82 \\
20 & Female & 21 & Technology company employee & Often & Often & General fortune prediction & Belief & Career continuity decision & Decision & 60 & 50 \\
\bottomrule
\end{tabular}
\end{table*}

\clearpage
\begin{table}[H]
\centering
\footnotesize
\setlength{\tabcolsep}{4pt}
\caption{Prompt structures used in LLM-based fortune-telling prompts, identified through a content analysis of 1,045 social media posts.}
\label{tab:rq1_prompt_structure}
\begin{tabular}{p{\dimexpr0.10\textwidth\relax}
                >{\centering\arraybackslash}p{\dimexpr0.14\textwidth\relax}
                p{\dimexpr0.60\textwidth\relax}
                >{\centering\arraybackslash}p{\dimexpr0.12\textwidth\relax}}
\toprule
\multicolumn{1}{c}{\textbf{Categories}} & 
\multicolumn{1}{c}{\textbf{Codes}} & 
\multicolumn{1}{c}{\textbf{Sample}} & 
\multicolumn{1}{c}{\textbf{Posts}} \\
\midrule

\multirow{9}{*}{\parbox[c]{1.6cm}{\centering Prompt Structure \cite{google_prompt_generative_AI_2025,liu2023pre}}} 
& Character & ``You are a professional researcher of traditional Chinese numerology.'' & 137 (13.11\%) \\
\cline{2-4}
& Knowledge & ``You are familiar with the Book of Fortune and Poverty, Divine Peak Examination, etc...'' & 159 (15.22\%) \\
\cline{2-4}
& Rule & ``For a male born in a Yang year, arrange in order; for a female born in a Yin year, arrange in order. For reverse order, use the month pillar as the reference.'' & 120 (11.48\%) \\
\cline{2-4}
& Personal Information & ``I was born on [year] [month] [day] at [hour]:[minute], female, in [place of birth].'' & 219 (20.96\%) \\
\cline{2-4}
& Task & ``Please analyze my birth chart as thoroughly as possible.'' & 228 (21.82\%) \\
\cline{2-4}
& Output Format & ``Please provide the output in the following format: 1) Overall fortune (brief summary) 2) Career and studies 3) Love and relationships 4) Wealth and finances...'' & 49 (4.69\%) \\
\cline{2-4}
& Tone & ``Please explain it in simple and easy-to-understand language.'' & 26 (2.49\%) \\
\cline{2-4}
& Required Explanation & ``Please explain the basis for your judgment.'' & 47 (4.50\%) \\
\cline{2-4}
& \textcolor{gray}{None} & \textcolor{gray}{-} & \textcolor{gray}{801 (76.65\%)} \\
\bottomrule
\end{tabular}
\end{table}

\clearpage
\appendix

\end{document}